\documentclass[]{aastex631}

\def\eqref#1{(\ref{eq:#1})}

\begin{document}

\title{Climatic Effects of Ocean Salinity on M dwarf Exoplanets}

\author{Kyle Batra} 
\affiliation{Department of Earth, Atmospheric, and Planetary Science, Purdue University, West Lafayette, IN, USA}
\affiliation{NASA Network for Ocean Worlds Exo-oceanography Team}
\affiliation{Alternative Earths NASA ICAR Team}
\author{Stephanie Olson}
\affiliation{Department of Earth, Atmospheric, and Planetary Science, Purdue University, West Lafayette, IN, USA}
\affiliation{NASA Network for Ocean Worlds Exo-oceanography Team}
\affiliation{Alternative Earths NASA ICAR Team}

\begin{abstract}
Ocean salinity is known to dramatically affect the climates of Earth-like planets orbiting Sun-like stars, with high salinity leading to less ice and higher surface temperature. However, how ocean composition impacts climate under different conditions, such as around different types of stars or at different positions within the habitable zone, has not been investigated. We used ROCKE-3D, an ocean-atmosphere general circulation model, to simulate how planetary climate responds to ocean salinities for planets with G-star vs. M-dwarf hosts at several stellar fluxes. We find that increasing ocean salinity from 20 to 100 g/kg in our model results in non-linear ice reduction and warming on G-star planets, sometimes causing abrupt transitions to different climate states. Conversely, sea ice on M-dwarf planets responds more gradually and linearly to increasing salinity. Moreover, reductions in sea ice on M-dwarf planets are not accompanied by significant surface warming as on G-star planets. High salinity can modestly bolster the resilience of M-dwarf planets against snowball glaciation and allow these planets to retain surface liquid water further from their host star, but the effects are muted compared to G-star planets that experience snowball bifurcation and climate hysteresis due to the ice-albedo feedback.

\end{abstract}

\section{Introduction} \label{sec:intro}

Earth's climate and habitability are directly influenced by its ocean \citep{cullum_importance_2016, ferreira_climate_2011, olson_effect_2022}. The same will also be true for exoplanets in their star's habitable zone, the circumstellar region where liquid water oceans are possible at a planet’s surface assuming that atmospheric pCO$_{2}$ is modulated by a functional silicate-weathering thermostat \citep{kasting_habitable_1993, kasting_remote_2014}. However, the role of exo-oceans in planetary climate may differ from Earth because exo-oceans may have diverse physiochemical properties \citep{olson_oceanographic_2020}. Ocean salinity is a particularly important property because salt lowers the freezing point of seawater \citep{fofonoff_algorithms_1983}, and it impacts ocean circulation through its effects on the density of sea water \citep{cullum_importance_2016}. Specifically, higher salinity can strengthen overturning circulation, increasing ocean heat transport to high latitudes \citep{cael_oceans_2017}. Both of these effects tend to decrease the spatial extent of sea ice in salty oceans \citep{olson_effect_2022}. 

Earth's present-day ocean has a salinity of 35 g/kg on global average and freezes at $-1.9 ^{\circ}$ C, but natural waters on Earth have a large range of salinities and freezing points. Many lakes have fresh water with negligible salt, while other bodies are hypersaline (100+ g/kg) such as the Dead Sea, which has a salinity in the range of 250 g/kg and a freezing point close to -21 $^{\circ}$C \citep{mor_effect_2018}. Moreover, ocean salinity has likely varied throughout Earth's history as the mass of dissolved salt and the volume of the surface water inventory has evolved \citep{holland_chemical_1984, knauth_salinity_1998, knauth_temperature_2005, hay_evaporites_2006, marty_salinity_2018, albarede_chemical_2020}. We thus expect to encounter exoplanets with ocean salinities different from our own, and it is important to anticipate how these variations may affect habitability under a variety of planetary scenarios, such as planets with different stellar hosts or positions in the habitable zone than Earth. 

M dwarfs are small stars with low temperatures and luminosities compared to Sun-like (G) stars. M dwarfs comprise approximately 80\% of main-sequence stars \citep{dressing_occurrence_2015}, and their relatively small size means that their habitable zones are closer in. This means that M-dwarf systems are more favorable for characterizing potentially habitable exoplanets in the near term with transit spectroscopy than G-star systems. Nonetheless, the habitability of terrestrial planets in M-dwarf systems remains uncertain due to many features that differ from their G-star planet counterparts \citep[reviewed by][]{shields_habitability_2016}. 

Encouragingly, M-dwarf planets are more resilient against global ``snowball" glaciation than planets with Sun-like G-star hosts because the ice-albedo feedback that was responsible for multiple episodes of global glaciation during Earth's habitable history is weakened \citep{joshi_suppression_2012, shields_effect_2013, von_paris_dependence_2013, checlair_no_2019}. The ice-albedo feedback arises on Earth because snow and ice are highly reflective at the visible wavelengths that dominate the Sun's spectrum \citep{brandt_surface_2005, nicolaus_modern_2010}. This means that increases in ice extent increase planetary albedo, reducing the absorption of stellar energy, cooling the surface, and promoting additional ice formation. Conversely, reductions in sea ice cover that expose low-albedo ocean can be amplified by increased absorption of stellar energy. This ice-albedo feedback leads to a snowball bifurcation in which small changes in radiative forcing, perhaps due to changing pCO$_2$, can induce rapid transitions between distinct climate states \citep{sellers_global_1969, budyko_effect_1969,ebert_intermediate_1993}. The ice-albedo feedback also yields a bistable climate system on G-star planets where the same instellation and greenhouse effect are compatible with dramatically differing climate states due to the important role of surface albedo in planetary energy balance \citep{ferreira_climate_2011, zhu_multiple_2023}. However, snow and ice have lower reflectivity at the IR wavelengths that dominate M-dwarf spectra, muting the impact of changing ice cover on the energy budget and climate of M-dwarf planets \citep{joshi_suppression_2012, shields_effect_2013, von_paris_dependence_2013, checlair_no_2019}.

Ocean salinity is ultimately able to have a major influence on the climate of Earth because its effects on ocean heat transport and freezing temperature jointly modulate the extent of sea ice, and these effects may be amplified by the ice-albedo feedback \citep{olson_effect_2022}. The magnitude of these effects and the importance of ocean salinity for planetary climate will thus differ on M-dwarf planets with less reflective ice \citep{joshi_suppression_2012, shields_effect_2013}. Atmospheric absorption of incoming IR radiation by CO$_2$, H$_2$O, and other gases also mutes climate sensitivity to surface ice and albedo \citep{von_paris_dependence_2013, shields_spectrum-driven_2014}, likely further modifying the impact of ocean salinity on M-dwarf planets compared to G-star planets. This absorption of incident IR radiation on M-dwarf planets ultimately occurs at the expense of surface absorption \citep{eager-nash_implications_2020}, likely modifying the balance of atmospheric vs. oceanic heat transport, which impact sea ice differently \citep{aylmer_impacts_2020}. However, how ocean salinity impacts planetary climate around different types of stars or at different stellar fluxes has not been thoroughly investigated. In this paper, we explore how stellar spectrum and ocean salinity jointly influence climate across a range of stellar fluxes (i.e., positions within the habitable zone) using a general circulation model (GCM) called ROCKE-3D.

\section{Methods} \label{sec:methods}
\subsection{Model Description}
We use the Planet 1.0 release of the Resolving Orbital and Climate Keys of Earth and Extraterrestrial Environments with Dynamics planetary GCM (ROCKE-3D) developed by NASA GISS \citep{way_resolving_2017}. ROCKE-3D is a fully coupled ocean-atmosphere GCM, modified from the GISS ModelE2-R Earth GCM \citep{schmidt_configuration_2014} to allow the simulation of a greater diversity planetary scenarios. ROCKE-3D has the flexibility to explore exoplanets with a variety of spin-orbit configurations \citep{way_climates_2018, jansen_climates_2019, del_genio_climates_2019, colose_effects_2021, he_climate_2022}, differing ocean and continent distributions \citep{salazar_effect_2020}, and stellar flux \citep[e.g.,][]{way_climates_2018, olson_effect_2022}. ROCKE-3D uses the versatile SOCRATES radiation scheme \citep{edwards_studies_1996,edwards_efficient_1996}, which also enables the user to swap out the Sun for other stellar hosts, including M dwarfs \citep{checlair_no_2019, del_genio_habitable_2019, salazar_effect_2020, fauchez_trappist-1_2020, colose_effects_2021}.

\subsection{Experiment Setup}
We simulate steady-state climates for planets orbiting the Sun (a G-star) and Proxima Centauri (an M dwarf; \citealt{meadows_habitability_2018}. We consider two luminosity scenarios for both stars: (1) the flux received by Earth at present-day (1,360.67 W/m$^2$ or 1.0 S$_\earth$); and (2) the flux received by Earth during the Archean Eon (1,108 W/m$^2$ or 0.814 S$_\earth$) \citep{gough_solar_1981}. We additionally use a lower instellation of 1,020.5 W/m$^2$ (0.75 S$_\earth$) for our M-dwarf simulations, as these planets tend to have warmer climates than their G-star counterparts \citep{shields_habitability_2016}. 

We explore a range of ocean salinities in our simulations that is broadly inclusive of natural waters on present-day Earth and estimates of the salinity of Earth's ocean through time \citep{knauth_temperature_2005, hay_evaporites_2006, marty_salinity_2018, albarede_chemical_2020}. For each star, we consider five global-mean ocean salinities: 20, 35 (present-day Earth), 50, 70, and 100 g/kg. ROCKE-3D is spun-up from an initially homogeneous ocean, but at steady state, salinity varies spatially in the ocean due to local precipitation/evaporation and sea ice formation. 

Each of our simulations has an atmosphere with 40 vertical layers up to 0.1 hPa, and they have a latitude-longitude resolution of $4^{\circ}\times 5^{\circ}$. Our simulations assume an Earth-like planet (mass, radius, and gravity) with a 1 bar N$_2$-dominated atmosphere with trace pre-industrial pCO$_2$, but no oxygen or ozone. All of our model scenarios assume an Earth-like rotation period, but we have disabled seasonality by imposing an eccentricity and obliquity of $0^{\circ}$. Finally, we use Earth's present-day continental configuration, assuming vegetation-free land surfaces consisting of a 50/50 blend of bright soil and dark soil. 

We first use ROCKE-3D's fully coupled configuration with a dynamic ocean to determine how ocean salinity affects planetary climate through a combination of changes to ocean heat transport, sea ice dynamics, and freezing point depression with different stellar spectra. This configuration includes a 10-layer, flat-bottomed ``bathtub" ocean that is 1,360 m deep \citep{way_climates_2018}. This simplification allows for simulations to come to steady state much faster than an a deeper ocean more similar to Earth's present-day ocean. It also limits computational expense while still being able to reproduce major components of Earth's ocean circulation, including the meridional overturning circulation. 

We then disable the dynamic ocean module and implement a slab ocean configuration with a depth of 50 m. The slab ocean still exchanges heat with the atmosphere but it has no interactive heat transport (zero ``q-flux''). Although salinity can be easily modified in netCDF input files for simulations with dynamic oceans \citep{del_genio_habitable_2019}, our simulations with slab oceans do not read salinity from standard input files. In these scenarios, we change ocean salinity by directly modifying ROCKE-3D's slab ocean module, in which salinity is defined as a constant that is then used to calculate freezing temperature \citep{olson_effect_2022}. These simulations without ocean and sea ice dynamics allow us to isolate the effects of freezing point depression vs. all dynamical effects for a given star type and instellation. 

We begin each experiment from an ice-free, warm state and run the model to steady state. We define steady state as the achievement of net global radiation $<$0.2 Wm$^{-2}$ and stabilization of surface temperature and sea ice cover \citep{way_resolving_2017}. Model scenarios may require 1,000+ years to reach steady state for warmer climates with dynamic oceans, but snowball states and slab ocean scenarios often require only a few hundred years to reach steady state. The data presented here are the average of the last decade of the simulation after the planetary climate has reached steady-state conditions, regardless of the duration of the simulation. These experiments are summarized in Table \ref{tab_var1}.

\begin{table}
\caption{Key Variables in Model Configuration}
\begin{center}
\begin{tabular}{l l l}
\hline
Parameter & Values & References \\
\hline
Instellation & 1.0, 0.814, 0.75 $S_\earth$ & \cite{gough_solar_1981} \\
Ocean Salinity & 20, 35, 50, 70, 100 g/kg & \\
Planetary Obliquity & $0^{\circ}$ & \\
Star Spectrum & Sun, Proxima Centauri & \cite{meadows_habitability_2018}\\
Ocean Depth & 1,360 m (dynamic bathtub), 50 m (slab) & \\
\hline
\end{tabular}
\item{Parameters not specified are fixed at present-day Earth values.}
\label{tab_var1}
\end{center}
\end{table}

\section{Results} \label{sec:results}

\begin{figure}
\plotone{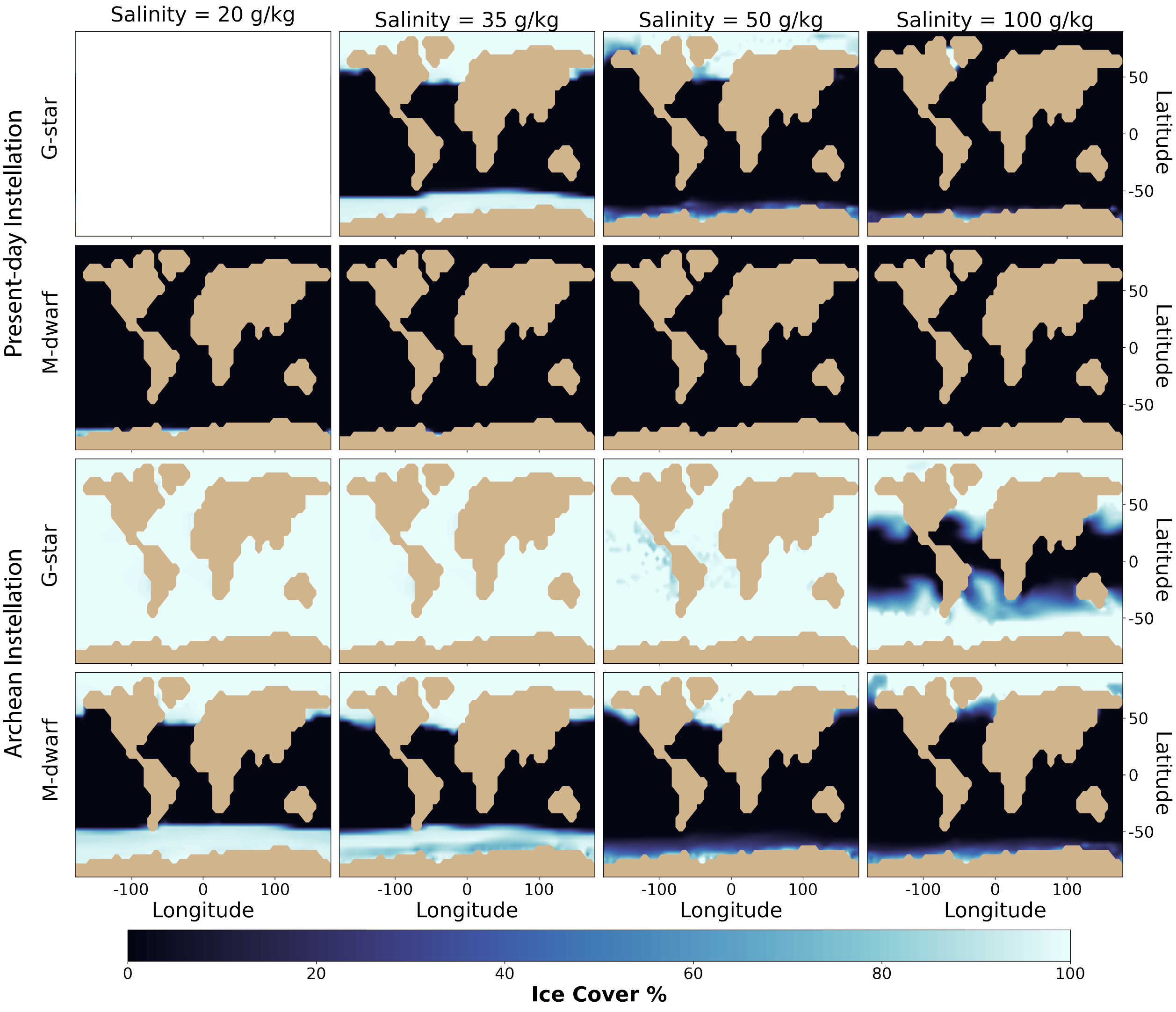}
\caption{Sea ice maps for G-star and M-dwarf exoplanets at present-day and Archean instellation with salinities from 20 to 100 g/kg, increasing from left to right. Each simulation has a 1 bar N$_2$-dominated atmosphere with trace CO$_2$, a planetary obliquity of $0^{\circ}$, a 24-hour day, and 365-day year.
\label{fig:icearch}}
\end{figure}

\begin{figure}[ht!]
\plotone{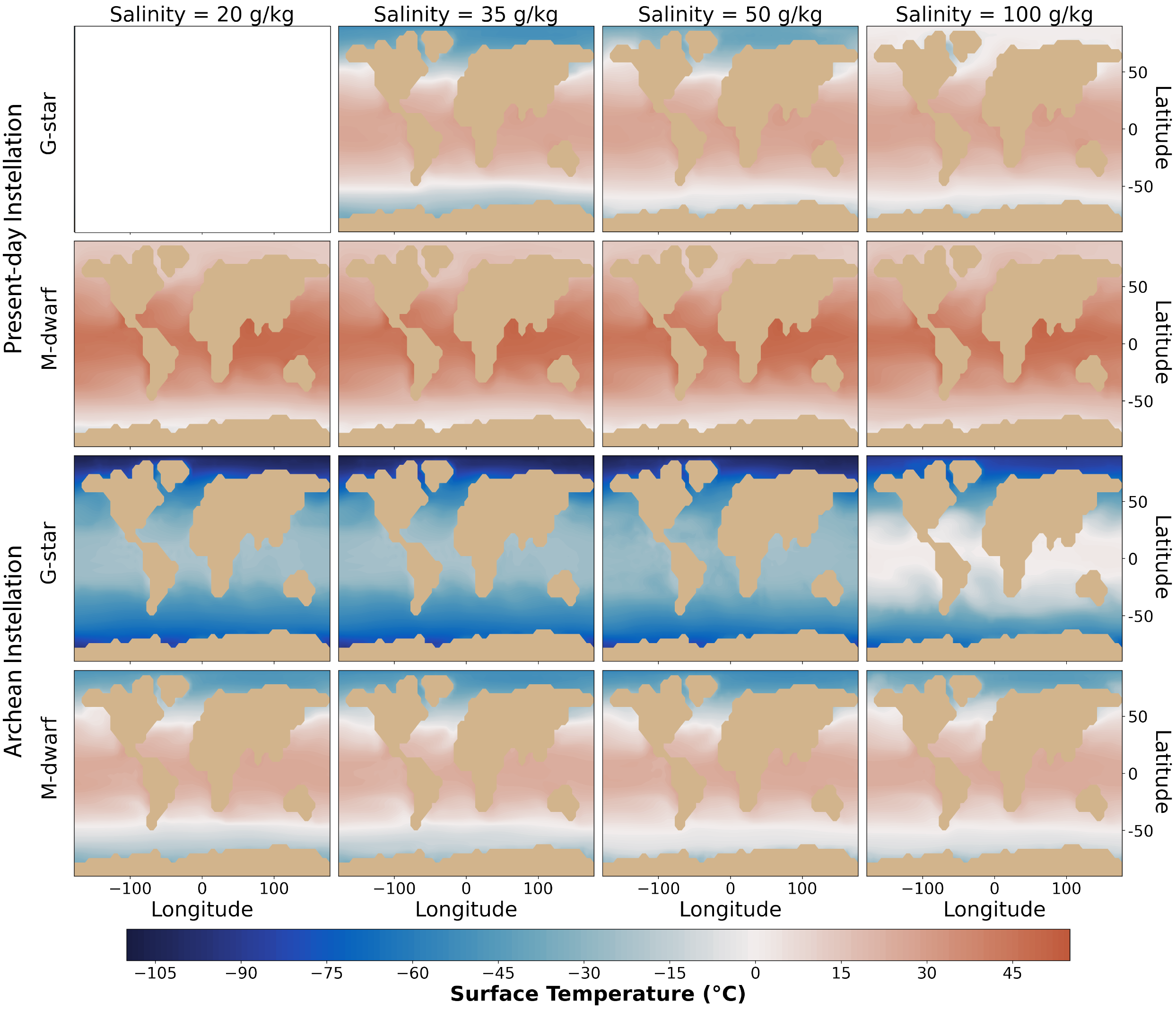}
\caption{Surface temperature maps for G-star and M-dwarf planets at present-day and Archean instellation with salinities from 20 to 100 g/kg, increasing from left to right. Each simulation has a 1 bar N$_2$-dominated atmosphere with trace CO$_2$, a planetary obliquity of $0^{\circ}$, a 24-hour day, and 365-day year.
\label{fig:temparch}}
\end{figure}

In our G-star simulations with present-day Earth instellation, sea ice was limited to high latitudes and decreased from 19.5\% to 3.5\% coverage with increasing ocean salinity from present-day values (35 g/kg) to our highest salinity scenario (100 g/kg). The average surface temperature of these G-star planets increased from 8 $^{\circ}$C with present-day Earth ocean salinity to 14 $^{\circ}$C in our highest salinity scenario (Figures \ref{fig:icearch} \& \ref{fig:temparch}). We note that our most Earth-like planetary scenario is somewhat cooler than present-day Earth because our simulations have zero obliquity, reducing illumination of the poles in our experiments relative to the actual Earth. We also find that the equator-to-pole temperature contrast is smaller at higher ocean salinities. Between our present-day and highest salinity scenario, the magnitude of the equator-to-pole contrast is reduced by $\sim$29 $^{\circ}$C, indicating that adding salt to the ocean preferentially warms the poles. 

Our simulated planets orbiting M-dwarfs have significantly less ice and higher surface temperatures than their G-star counterparts for a given stellar flux, in agreement with previous work \citep[reviwed by][]{shields_habitability_2016}. Sea ice cover was limited for all salinities and negligibly decreased between our present-day and highest ocean salinity scenarios for M-dwarf planets receiving present-day Earth instellation (Figure \ref{fig:icearch}). The average surface temperature of these M-dwarf planets increased only slightly from 31.5 to 32 $^{\circ}$C between these ocean salinity scenarios (Figure \ref{fig:temparch}). Meanwhile, the equator-to-pole temperature contrast by $\sim$2 $^{\circ}$C, which is small compared to our equivalent G-star simulations. The impact of salinity on climate is thus strongly dampened for this stellar scenario. 

The average sea ice cover of G-star planets that have the same present-day Earth atmosphere but receive lower Archean-like instellation decreased from $\sim$100\% with present-day Earth salinity to 45\% with our highest salinity ocean (Figure \ref{fig:icearch}). The average surface temperature in these same scenarios increased from -41 $^{\circ}$C to -21 $^{\circ}$C (Figure \ref{fig:temparch}), and the equator-to-pole temperature contrast decreased by $\sim$38 $^{\circ}$C. Nonetheless, these planets were all severely, if not globally, glaciated. Modestly increasing ocean salinity from 70 g/kg to 100 g/kg prevented global glaciation, reducing planetary albedo and significantly increasing surface temperature. These two ocean salinity scenarios thus produce distinct climate states. 

In contrast, the same Archean-like instellation did not lead to extensive glaciation on M-dwarf planets (Figures \ref{fig:icearch} \& \ref{fig:temparch}). We find that the low-to-mid latitudes of every M-dwarf planet with Archean-like instellation still had temperatures far above freezing and ice extent was restricted to latitudes north and south of 50$^{\circ}$. The sea ice cover of M-dwarf planets receiving Archean-like instellation decreased from 23.2\% to 7.9\% with increasing ocean salinity.  The average surface temperature increased modestly from 5 to 8 $^{\circ}$C between our present-day and highest salinity simulations while the equator-to-pole temperature contrast decreased by only $\sim$1.5 $^{\circ}$C. Moreover, whereas we saw abrupt changes in climate state between some of our G-star planets with Archean instellation and different ocean salinities, changes in ice cover and temperature were more gradual and continuous in the equivalent M-dwarf simulations. 

M-dwarf planets with even lower instellation (0.75 S$_\earth$) still had less sea ice than the G-star planets with higher, Archean-like instellation (0.814 S$_\earth$) at equivalent salinities. The average ice cover of these M-dwarf planets receiving low instellation decreased from 38.3\% to 14.8\% between our present-day and highest ocean salinities. Meanwhile, the average surface temperatures of the M-dwarf planets receiving low instellation increased linearly from -5.7 to -1.9 $^{\circ}$C for present-day Earth's ocean salinity vs. our highest ocean salinity scenario. The reduction in equator-to-pole temperature contrast between these salinity scenarios was $\sim$2.5 $^{\circ}$C.

\begin{figure}[ht!]
\plotone{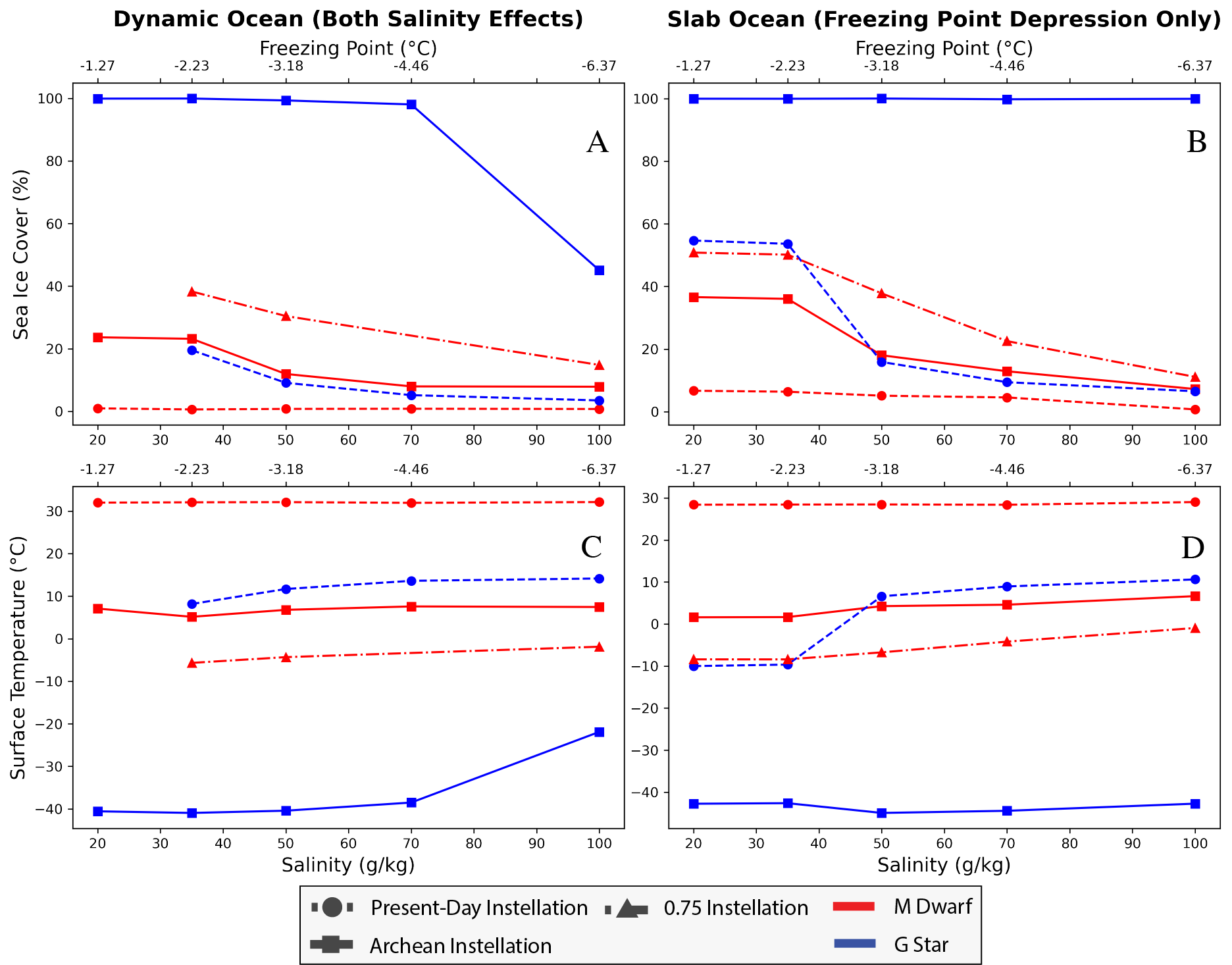}
\caption{Global sea ice cover (A, B) and average surface temperature (C, D) of G-star vs. M-dwarf planets as a function of ocean salinity (bottom axis), which determines the freezing point of seawater assuming 100\% dissociation of pure NaCl salt (top axis). The left column (subplots A, C) shows fully coupled simulations with a ``Dynamic Ocean", whereas the right column labeled ``Slab Ocean" (subplots B, D) presents results from simulations excluding all ocean and sea ice dynamics. Climate sensitivity in the Slab Ocean simulations arises only due to freezing point depression. Red lines denote M-dwarf (Proxima Centauri) simulations whereas blue lines represent G-star (Sun) simulations. The dashed line with circle symbols represent present-day Earth instellation (1.0 S$_\earth$), the solid line with square symbols depict Archean-like instellation (0.814 S$_\earth$), and the dot-dash line with triangle symbols show the lowest instellation (0.75 S$_\earth$; M-dwarf planets only).
\label{fig:TempIcePlot}}
\end{figure}

\begin{figure}[ht!]
\centering
\includegraphics[width=1.0\textwidth]{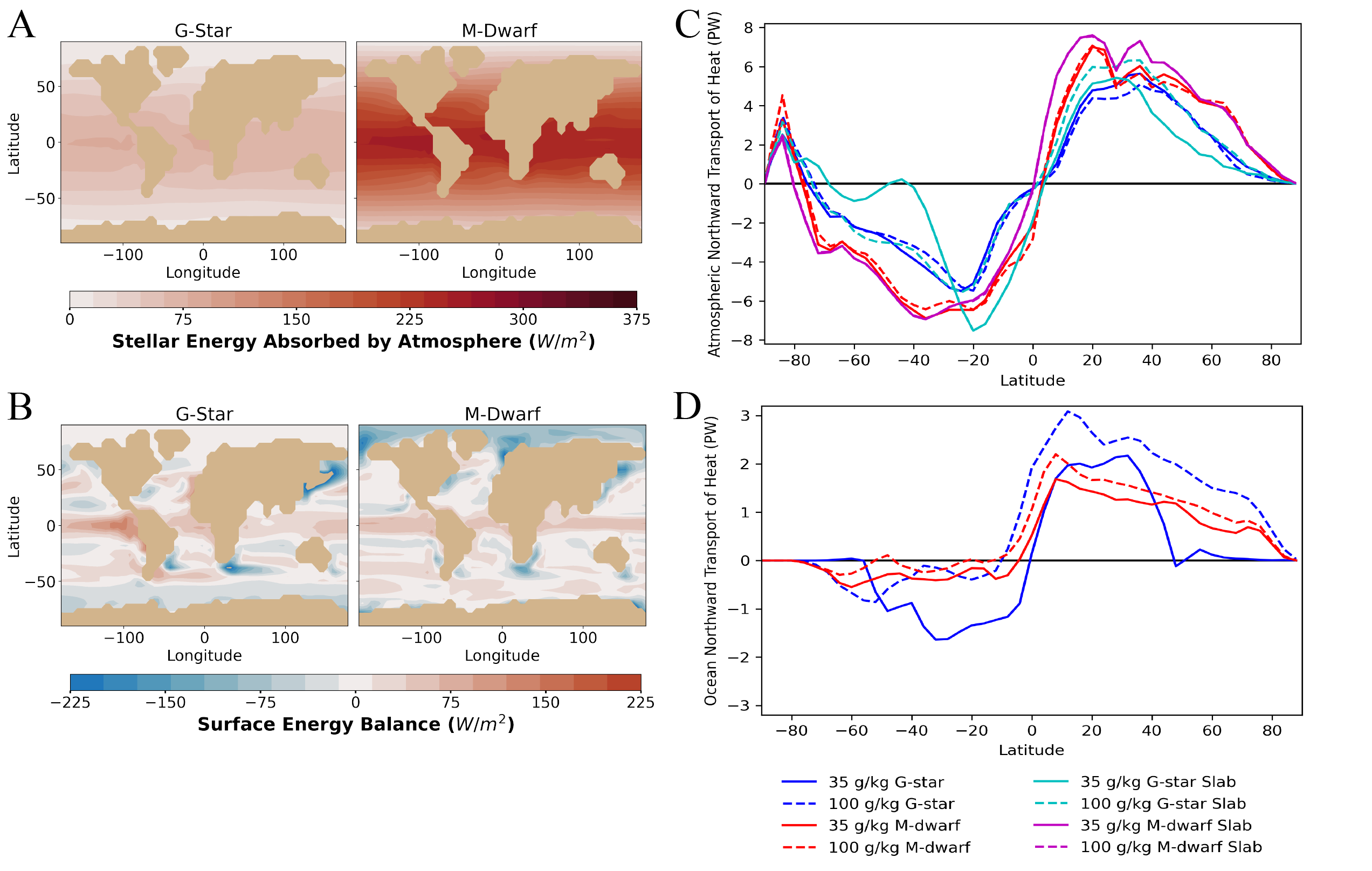}
\caption{Atmosphere and ocean heat transport and energy budget on G-star vs. M-dwarf planets. Panel A shows the absorption of incoming stellar radiation by the atmospheres of planets orbiting G-stars (left) and M-dwarfs (right) for simulations with Earth's present-day instellation and ocean salinity (35 g/kg). Panel B shows the net balance of energy in/out of the ocean system, inclusive of incident shortwave, incident longwave, sensible heat, and latent heat fluxes at the oceans surface for the same model scenarios. Panel C shows northward atmospheric heat transport for present-day instellation G-star and M-dwarf planets with ocean salinities of 35 g/kg and 100 g/kg with both a dynamic and slab ocean. Panel D then shows northward ocean heat transport for present-day instellation G-star and M-dwarf planets with ocean salinities 35 and 100 g/kg.
\label{fig:heat}}
\end{figure}

For both star types, poleward ocean heat transport increases while atmospheric heat transport decreases with salinity in the Northern Hemisphere (Figure \ref{fig:heat} C, D). However, poleward ocean heat transport generally decreases with increasing salinity in the Southern Hemisphere. For the G-star scenario shown in Figure \ref{fig:heat} D, the increase in ocean heat transport south of -60$^{\circ}$ may be a consequence of ice retreat rather than a driver of ice loss \cite{rose_role_2013}.

Overall, our M-dwarf simulations have reduced ocean heat transport relative to equivalent G-star simulations with dynamic oceans (Figure \ref{fig:heat} D). This difference arises because the atmospheres of M-dwarf planets absorb substantially more incident radiation (Figure \ref{fig:heat} A) and have greater atmospheric heat transport (Figure \ref{fig:heat} C), ultimately reducing energy incident upon and absorbed by the underlying ocean compared to G-star scenarios (Figure \ref{fig:heat} B). 

We then disabled ocean dynamics by replacing ROCKE-3D's dynamic ocean with a 50 m slab ocean without heat transport (zero q-flux). With this slab ocean, the G-star planets receiving present-day Earth instellation are unsurprisingly much icier and colder than equivalent simulations with a dynamic ocean (Figure \ref{fig:TempIcePlot} B, D), reflecting an important role for ocean dynamics in transporting heat to high latitudes and curtailing sea ice advance \citep{winton_climatic_2003, bitz_maintenance_2005, ferreira_climate_2011}. Some G-star planets with slab oceans have a distinct climate state from equivalent simulations with a dynamic ocean. On the other hand, the climate of M-dwarf planets is more similar between our simulations with dynamic versus slab oceans but with the same salinity (Figure \ref{fig:TempIcePlot}), suggesting that ocean dynamics play a lesser role in the climates of these worlds. For both star types, poleward atmospheric heat transport is greater in simulations with a slab ocean compared to those with a dynamic ocean, but atmospheric transport compensates for the lack of ocean heat transport more effectively for M-dwarf planets (Figure \ref{fig:heat} C). 
Overall, eliminating ocean dynamics did not strongly modify either the baseline ice cover with present-day Earth ocean salinity or its sensitivity to salinity at all instellations for M-dwarf planets, unlike G-star planets (Figure \ref{fig:TempIcePlot}). 

\section{Discussion} \label{sec:discussion}

Increasing ocean salinity reduces ice cover and warms the climates of planets orbiting both G-star and M-dwarf hosts, but these effects vary both qualitatively and quantitatively between star types. We explore these differences in subsection \ref{subsec:impact}, consider implications for habitability and exoplanet life detection in Section \ref{subsec:implications}, and discuss opportunities for future work in Section \ref{subsec:work}.

\newpage
\subsection{The Impact of Ocean Salinity on G vs M-dwarf Planets} \label{subsec:impact}

Ocean salinity affects both ocean dynamics (and thus ocean heat transport) and the freezing temperature of seawater. Higher ocean salinity can strengthen overturning circulation and ocean heat transport to high latitudes (Figure \ref{fig:heat} D), reducing equator-to-pole temperature contrasts and limiting sea ice extent \citep{ferreira_climate_2011}. Higher salinity also decreases the freezing temperature of seawater, directly inhibiting sea ice formation. On G-star planets these changes may be amplified by the ice-albedo feedback, and in some cases low vs. high salinity can produce distinct climate states with dramatically different ice cover and surface temperatures \citep{olson_effect_2022}. 
This is best illustrated in our highest salinity (100 g/kg) scenario in which our simulation with ocean dynamics avoided global glaciation at Archean-like instellation, but entered a snowball state when ocean dynamics were disabled and freezing point depression was salinity's sole warming effect. These results are in agreement with a previous study that explored the effects of salinity on the climate of present-day Earth, which found simulations which included ocean dynamics but excluded freezing point depression, were able to effectively reproduce the results of simulations that included both salinity effects, while simulations with only freezing point depression were relatively insensitive to changing ocean salinity \citep{olson_effect_2022}. The disparity between our G-star simulations with and without ocean dynamics and the inability of slab ocean simulations to approximate climate effects to ocean salinity illustrate the value of including dynamic oceans in simulations of G-star exoplanets. 

We find that the effect of ocean salinity on the climate of M-dwarf planets is muted compared to G-star planets. There are three key reasons. First, our simulations show that ocean heat transport is overall weaker on M-dwarf planets while atmospheric absorption of incident radiation strengthens atmospheric heat transport on M-dwarf planets compared to equivalent G-star planets receiving the same flux (Figure \ref{fig:heat}). However, atmospheric heat transport is not directly influenced by ocean salinity like ocean heat transport and sea ice is less sensitive to atmospheric heat transport than ocean heat transport \citep{aylmer_impacts_2020}, reducing the effects of salinity on climate. Second, although ocean heat transport does strengthen with increasing salinity on M-dwarf planets, this change is small compared to equivalent G-star planets. Third, incremental reductions in sea ice extent arising from freezing point depression with increasing salinity have a minimal effect on planetary albedo on M-dwarf planets. These changes thus do not affect surface temperature via the ice-albedo feedback as on G-star planets. 
Taken together, our simulations suggest that in many instances including a computationally expensive dynamic ocean will not affect first-order conclusions regarding the habitability of M-dwarf planets and that climate sensitivity to ocean salinity may be reasonably approximated by modifying the freezing point in simulations of M-dwarf planets with models lacking dynamic oceans such as ExoCAM or ExoPlaSim.

For both G-star and M-dwarf simulations, there is an asymmetry in the hemispheric response of ocean heat transport to increasing salinity (Figure \ref{fig:heat} D). Whereas higher salinity uniformly increases poleward ocean heat transport in the northern hemisphere, poleward ocean heat transport is overall weaker in the southern hemisphere (Figure \ref{fig:heat} D), consistent with previous work suggesting that AMOC strengthens while PMOC weakens as Earth's ocean becomes saltier \citep{cael_oceans_2017}.
However, AMOC and PMOC are a result of present-day Earth's continental configuration. In previous aquaplanet simulations lacking land, poleward heat transport increases with ocean salinity symmetrically in both hemispheres \citep{olson_effect_2022}. Although we expect higher ocean salinity to impact ocean heat transport regardless of continental configuration spatial details from our simulations cannot be generalized to all continental configurations. 

Like ocean heat transport, sea ice retreat is hemispherically asymmetric in our simulations. Surprisingly, sea ice in the Antarctic region is lost first and appears more responsive to changes in ocean salinity compared to sea ice in the Arctic despite ocean heat transport weakening in the Antarctic while strengthening in the Arctic. These results suggest that sea ice retreat in the Antarctic is largely controlled by freezing point depression. Arctic sea ice may be stabilized against increasing ocean heat transport and freezing point depression by compensating changes in atmospheric heat transport (Figure \ref{fig:heat}(C) and (D); \citep{aylmer_different_2022}. 
The relative importance of the freezing point depression compared to ocean heat transport for the influence of salinity on the climate of other G-star planets may vary spatially and will depend on the continental configuration, which shapes the large-scale ocean circulation patterns.

\subsection{Implications for Habitability and Life Detection} \label{subsec:implications}
\cite{olson_oceanographic_2020} raised concerns regarding how exo-ocean characteristics, like salinity, may impact planetary climate and habitability while being impossible to directly constrain remotely. Indeed, multiple distinct climates are possible for the same atmospheric pCO$_{2}$ and stellar flux depending on surface albedo and thus ocean salinity, which affects ice cover in our simulations. 

While the climatic impact of an unknown ocean salinity must be considered for G-star planets in the habitable zone \citep{cullum_importance_2016, olson_effect_2022}, we show that high vs. low ocean salinity does not result in distinct climate states on M-dwarf planets due to muted changes in ice extent from weaker ocean dynamics and ice-albedo feedback. On these planets, increasing salinity results in incremental changes in ice cover with minimal changes in surface temperature. Unlike G-star planets, uncertainty regarding ocean salinity on M-dwarf planets is therefore not likely to change first-order conclusions regarding planetary habitability based on orbital separation and spectroscopic pCO$_{2}$ constraints. This is encouraging news for the search for life on M-dwarf planets, but several additional questions on M-dwarf planetary habitability require future investigation including their spin-orbit regime \citep[e.g.,][]{merlis_atmospheric_2010, haqq-misra_demarcating_2018, colose_effects_2021}, planetary obliquity in multi-planet systems \citep[e.g.,][]{wang_effects_2016, kane_obliquity_2017, valente_tidal_2022}, and host star activity \citep[e.g.,][]{lammer_coronal_2007, luger_rodrigo_extreme_2015, bolmont_water_2017}. 

Beyond climate, ocean salinity will have additional impacts on marine habitats, the success of ocean life, and the detectability of biosignatures on both G-star and M-dwarf planets. For example, ocean circulation is a vital control on biological productivity in a planet's ocean because it shapes the spatial and temporal distribution of the chemical ingredients for life. Increasing salinity strengthens ocean overturning in our simulations of both G and M stars and may boost biological productivity by increasing the availability of essential nutrients that tend to accumulate in the deep ocean where they are unavailable for photosynthesis at the surface. This overturning also helps to transport biosignature gases produced at depth in the ocean to the surface, where they may enter the atmosphere and influence planetary spectra \citep{olson_oceanographic_2020}. Higher salinity also reduces gas solubility, further enhancing the escape of biological gases from salty oceans to the atmosphere--especially when combined with lower ice cover and higher temperatures. In other words, not only could planets with salty oceans remain habitable further from their hosts stars due to the climatic effects of salt, they may also support higher biological productivity and more efficient transfer of biosignature gases to the atmosphere. Both of these effects may make life easier to detect in spectra of planets with saltier oceans.

Although we focus on ocean habitats, ocean salinity also affects the habitability of land environments on G-star planets by modifying planetary albedo and energy balance. Salinity thus modulates the extent of continental ice sheets in addition to directly affecting sea ice. Because vegetation on land more directly influences remote observables than ocean life, saltier oceans may further amplify the detectability of life on G-star planets by increasing the land surface hospitable towards plant-like life. Conversely, the effects of ocean salinity on M-dwarf planets are largely limited to direct inhibition of sea ice formation, without significant warming or consequences for life on land. 

\subsection{Opportunities for Future Work} \label{subsec:work}

M-dwarf planets in the habitable zone are expected to be synchronously rotating (tidally locked) with permanent day and night sides due to their close proximity to the host star \citep{shields_habitability_2016}. In this scenario, sea ice is even less coupled to planetary albedo than in our simulations with Earth-like rotation because the ice on the night side would not interact with incoming radiation. Depending on the planet's rotation regime (rapid, Rhines, or slow), atmospheric and oceanic circulation and resulting heat transport to the night side and to high latitudes will vary \citep{merlis_atmospheric_2010, edson_atmospheric_2011, haqq-misra_geothermal_2015, noda_circulation_2017, haqq-misra_demarcating_2018}. Under different rotation and circulation regimes, the climate sensitivity to salinity may therefore differ. Considering how orbital scenarios such as synchronous rotation modify climate sensitivity to ocean salinity in tandem with stellar spectra is an opportunity for future study.

Future work could also explore the impact of different salt compositions on planetary climate. The freezing point depression calculation in ROCKE-3D assumes our exo-oceans are similar in composition to Earth's present-day ocean (i.e., NaCl dominated). However, other bodies of water on Earth have different salt compositions compared to the ocean, such as the magnesium chloride (MgCl$_2$) rich Dead Sea \citep{krumgalz_physico-chemical_1982}, and the major ions contributing to Earth's ocean salinity may have changed through time \citep{albarede_chemical_2020}. Both Mars and Europa have high concentrations or magnesium sulphate salts \citep{clark_salts_1981, zolotov_composition_2001}. These differences in ionic composition may shift freezing point by several degrees for equivalent salinities, with consequences for sea ice. Additionally, sea salt aerosol with different ionic composition may behave differently as cloud condensation nuclei (CCN) due to differing hygroscopicity \citep{prijith_relationship_2014}, potentially impacting planetary albedo for G-star and M-dwarf planets alike.

\section{Conclusions} \label{sec:conclusions}

Our ROCKE-3D simulations show that ocean salinity, which cannot be remotely characterized, plays an major role in planetary climate under some circumstances. In particular, ocean salinity can be the difference between globally glaciated states or climates with surface liquid water for habitable zone planets orbiting G-stars. These differences arise because higher salinity strengthens ocean heat transport and depresses the freezing point of seawater. Both effects lead to less sea ice. On G-star planets, the resulting changes in ice cover may then be amplified by the ice-albedo feedback. However, salinity has a more muted influence on the climates of M-dwarf planets. Ocean heat transport is both weaker and less sensitive to ocean salinity on M-dwarf planets. Moreover, the low reflectivity of ice at the IR wavelengths that dominate M-dwarf spectra means that changes in ice cover are not amplified by changes to planetary energy balance. The result is continuous and incremental reductions in sea ice cover with increasing salinity on M-dwarf planets without accompanying changes in surface temperature, in contrast to G-star planets that experience snowball bifurcation and nonlinear sensitivity to ocean salinity. While habitability outcomes may differ between G-star planets with differing ocean salinities, differences in ocean salinity are less likely to produce distinct climate states on M-dwarf planets. This is an encouraging result that suggests uncertainties regarding exo-ocean salinity are less of a concern for understanding the climates and habitability of M-dwarf planets compared to G-star planets. We also find that simulating the climates of M-dwarf planets with a slab ocean is less likely to impact first order conclusions regarding planetary habitability compared to G-star planets.
 
\section{Acknowledgements} \label{sec:acknowledgements}
We thank Chris Colose for a thoughtful and constructive review that significantly improved the manuscript. This work was supported by grants from the NASA Interdisciplinary Consortia for Astrobiology Research (ICAR), the NASA Habitable Worlds Program, and the Heising-Simons Foundation to SLO.


\bibliography{BatraMD1}{}
\bibliographystyle{aasjournal}

\end{document}